**Properties of (TiZrNbCu)$_{1-x}$Ni$_x$ Metallic Glasses**

Ignacio A. Figueroa[a], Ramir Ristić[b], Ahmed Kuršumović[c], Katica Biljaković[d], Damir Starešinić[d], Damir Pajić[e], György Remenyi[f], Emil Babić[e]

[a]Institute for Materials Research-UNAM, Ciudad Universitaria Coyoacan, C.P. 04510, Mexico D.F., Mexico
[b]Department of Physics, University of Osijek, Trg Ljudevita Gaja 6, HR-31000 Osijek, Croatia
[c]Department of Materials Science & Metallurgy, University of Cambridge, 27 Charles Babbage Road, Cambridge, CB 3 0FS, UK
[d]Institute of Physics, Bijenička cesta 46, P.O.Box 304, HR-10 001, Zagreb, Croatia
[e]Department of Physics, Faculty of Science, Bijenička cesta 32, HR-10002,Zagreb, Croatia
[f]Institut Neel, Universite Grenoble Alpes, F-38042, Grenoble, France

Corresponding author: Ramir Ristić ; e-mail ramir.ristic@fizika.unios.hr

Abstract

Recent studies (J. Alloys Compd. 695 (2017) 2661) of the electronic structure and properties of (TiZrNbCu)$_{1-x}$Ni$_x$ ($x \leq 0.25$) amorphous high entropy alloys (a-HEA) have been extended to $x = 0.5$ in order to compare behaviours of a-HEA and conventional Ni-base metallic glasses (MG). The amorphous state of all samples was verified by thermal analysis and X-ray diffraction (XRD). XRD indicated a probable change in local atomic arrangements, i.e. short-range-order (SRO) for $x \geq 0.35$. Simultaneously, thermal parameters, such as the first crystallization temperature $T_x$ and the liquidus temperature showed a tendency to saturate for $x \geq 0.35$. The same tendency also appeared in the magnetic susceptibility $\chi_{exp}$ and the linear term in the low temperature specific heat $\gamma$. The Debye temperatures and Young´s moduli also tend to saturate for $x \geq 0.35$. These unusual changes in SRO and all properties within the amorphous phase seem correlated with the change of valence electron number (*VEC*) on increasing $x$.

1. Introduction

High entropy alloys (HEA) are new types of alloys based on multiple principal alloying elements in near equi-atomic ratios [1-4]. This alloy design enables research and probable exploitation of a large number of completely new alloys with structures and properties which can hardly be anticipated [5,6]. As result, the research of HEAs [3] resulted in several hundreds of research papers [7], several reviews of the literature [6-13] and a book [14] in a short time.

A large research effort invested into the studies of phases, microstructures and mechanical properties of HEAs resulted in a wealth of data confirming their conceptual and technological importance [5-14]. In particular, HEAs with ultrahigh strength and fracture toughness, outstanding mechanical properties at high temperature, high fatigue, wear, corrosion and irradiation resistance were recently developed [6,7,11-14]. A large asset of HEAs is the simplicity of tuning their properties by adjusting their composition and/or phase content. Until recently amorphous (a-) HEAs received relatively little attention [15], in spite of the fact that they were the first applications of a new alloy design [1,2] and are of crucial importance for the understanding of some features specific to disorder, such as the boson peak. The overlap between a-HEAs and multiphase alloys in a plot of mixing enthalpies ($\Delta H_{mix}$) vs. the relative atomic size mismatch ($\delta$) is almost complete [13], thus the prediction of the formation of high-entropy bulk metallic glasses (HE-BMG) is probably even more difficult than that in conventional binary and multicomponent metallic glasses. This is probably one reason that in spite of an early start [1,2], the progress in research of a-HEAs was much slower than that of crystalline ones and that critical thickness of HE-BMGs is generally lower than that of conventional BMGs. However, more recently the situation started to change and several comprehensive studies showing the conceptual and technological relevance of a-HEAs appeared (e.g., [16,17]). In particular, recent study of formation, thermal stability and mechanical properties of ($Fe_{0.25}Co_{0.25}Ni_{0.25}Cr_{0.125}Mo_{0.125}$)$_{100-x}$B$_x$ melt-spun metallic glasses revealed that their glass forming range extends down to 11 at %B which is well below of that necessary for the glass formation in corresponding binary and ternary alloys [16]. Simultaneously, both the crystallization temperatures and hardness of these a-HEAs increase linearly with x which seems to indicate that, like in Ni-(B,Si,P) metallic glasses [18] chemical bonding between the transition metal(s) and metalloid(s) govern their strength and stability. Further, the alloys with x ≥ 22 at% B crystallize directly into an intermetallic compound ($M_{23}B_6$), thus experience direct transition from disordered to ordered state which is of great conceptual importance.

However, a conceptual understanding of both a- and crystalline (c-) single phase HEAs is still quite limited as seen from the core effects [7] and an excessive use of rule-of-mixtures (RoM) for the calculation and/or prediction of their properties (e.g. Refs. [5,12,15,19]). This is partially due to lack of insight into their electronic band structures (ES) which in metallic systems determine almost all properties (e.g. Ref. [20]) of a given system. Indeed, until recently [21] there were just three measurements of properties directly related to ES [19,22,23] and few calculations of ES and selected properties for single phase c-HEAs (e.g. Ref. [24]) and no such reports for a-HEAs.

Very recently we reported the first comprehensive study of the structure and morphology, ES and selected properties related to ES and interatomic bonding in (TiZrNbCu)$_{1-x}$Ni$_x$ ($x \leq 0.25$) a-HEAs [21]. This system was selected because of a previous report [25] that depending on preparation and/or processing conditions, alloys with $x \leq 0.15$ can form either c-(body centered cubic, bcc) or a-HEA. Although our analysis of XRD patterns of our a-HEAs indicated a bcc-like atomic arrangement [21], so far we have not been able to prepare single phase crystalline samples of our alloys. (This result is consistent with the analysis of parameters related to phase selection in HEAs [21,26] which indicate formation of intermediate phases in all our alloys.) The low temperature specific heat (LTSH) showed that the electronic density of states at the Fermi level ($E_F$), $N(E_F)$, decreases with increasing $x$, whereas the Debye temperature $\theta_D$ increases with $x$ [21]. This is similar to what is observed in binary and ternary amorphous alloys of early (TE) and late (TL) transition metals [27,28] and indicates that $N(E_F)$ is dominated by the d-electrons of TE. (Recent photoemission spectroscopy (PES) study of our alloys [29] confirmed that as in a-Zr-Ni-Cu alloys [30] d-states of Ni and Cu are well below $E_F$ and $N(E_F)$ is dominated by the d-states of TEs.) Like $\theta_D$, the Young´s moduli ($E$), the first crystallization temperatures ($T_x$) and liquidus temperatures ($T_L$) increased with increasing $x$ indicating enhanced inter-atomic bonding on addition of TL (similar to that observed in a-TE-TL alloys [27,31]).

Here, we extended a previous study [21] to $x = 0.5$ in order to compare the behaviours of a-HEAs and conventional TL-base metallic glasses (MG) from the same alloy system. New results for alloys with $x > 0.25$ seem to indicate a change in local atomic structure from bcc-like to possibly fcc-like which occurs for the average valence electron number (defined as $VEC = \sum_{i=1}^{n} c_i (VEC)_i$, where c$_i$ and (VEC)$_i$, are the atomic fraction and valence electron number for the i-th element, respectively) $VEC > 7$ (similar to what is observed in c-HEAs [13]). This affects the variation of all studied properties with $x$, which became very weak for $x \geq 0.35$, thus distinctly different from those in binary a-TE-TL alloys [27,28,31]) with similar TL content.

2. Experimental

The ingots of six alloys in the (TiZrNbCu)$_{1-x}$Ni$_x$ system with $x = 0.125, 0.15, 0.20, 0.25, 0.35$ and $0.5$ were prepared from high purity elements ($\geq 99.98$ %) by arc melting in high purity argon in the presence of a Ti getter. The ingots were flipped and re-melted five times in order to ensure good mixing of components. Ribbons with thickness of about 20 μm were fabricated by melt spinning molten alloys on the surface of a copper roller rotating at a speed of 25 m/s in a pure He atmosphere [21]. Casting with controlled parameters resulted in ribbons with closely similar cross-sections (~2 x 0.02 mm$^2$) and thus with amorphous phases having a similar degree of quenched-in disorder. The as-cast samples were investigated by: (1) XRD using a Bruker Advance powder diffractometer with a CuK$_\alpha$ source, (2) scanning electron microscopy (SEM) using a JEOL ISM7600F microscope with energy dispersive spectrometry (EDS) capability, and (3) differential thermal analysis (DTA) and differential scanning calorimetry (DSC) using a TA instrument Thermal Analysis-DSC-TGA. Thermal measurement were performed with a ramp rate of 20 K/min up to 1550 K. The DTA equipment is regularly calibrated using strontium carbonate and gold standards. This procedure keeps the uncertainty in temperatures derived from DSC-DTA measurements to within ± 5 K. As-cast ribbons were also used for measurements of LTSH, magnetic susceptibility and mass density D [21]. LTSH measurements were performed in the temperature range 1.8-300 K using a Physical Property Measurement System (PPMS), Model 6000 from Quantum Design Inc. [21,32]. The magnetic susceptibility was measured with a Quantum Design magnetometer, MPMS5, in a magnetic field $B \leq$ 5.5 T and temperature range 5-300 K [20,21,27]. Since the magnetic susceptibility of all samples showed a very weak dependence on temperature (as is usual in metallic glasses and compounds of early and late transition metals [20,21,27,33]), in the following analysis we will use their room temperature values. The Young´s modulus, $E$, which was calculated from the relationship $E = Dv^2$, where $v$ is the velocity of ultrasonic waves along the ribbon, was measured both on as-cast ribbons and the same ribbons relaxed for a short time close to the glass transition temperature of a given alloy [21,27].

3. Results and discussion

XRD patterns in Fig. 1. show that all our samples are fully amorphous with a characteristic broad first peak around $2\theta \approx 40^0$. Thus, the glass forming range (GFR) in (TiZrNbCu)$_{1-x}$Ni$_x$ alloys seems quite wide, extending at least from $x = 0.125$ to $x = 0.50$. We note that the first maxima in the XRD traces in Fig. 1. shift with increasing $x$ to larger values of $2\theta$ as could be expected due to small size of Ni atom. From the modulus of the scattering vector, $k_p$, corresponding to the first maximum of the diffraction patterns in Fig. 1. we calculated the average nearest-neighbour distances, $d$ [21,27,33] for all our amorphous alloys and assuming a bcc-like local atomic structure [21] we estimated the corresponding lattice parameters, $a=2d/3^{0.5}$. (Although approximate, this method previously provided fairly accurate estimates for the average atomic volumes of some Zr-Cu and Hf-Cu metallic glasses [27,33].) These values of $a$ are plotted vs. $x$ in Fig. 2. For comparison the theoretical lattice parameters calculated from the lattice parameters for the bcc phases of constituents by assuming the validity of the Vegard´s law are also shown in Fig. 2.. As noted earlier [21] for $x \leq 0.2$ the experimental values of $a$ agree quite well with those calculated using the Vegard´s law. (The experimental values are however slightly larger than corresponding theoretical values which may arise from the less dense atomic packing in the amorphous state [21].) This result is consistent with the appearance of a dominant bcc phase in bulk crystalline $(TiZrNbCu)_{1-x}Ni_x$ samples with $x \leq 0.2$ [25] and is also consistent with the *VEC* criterion [13] for the selection of crystal structure in c-HEAs: namely the bcc structure is stable for $VEC \leq 6.8$.

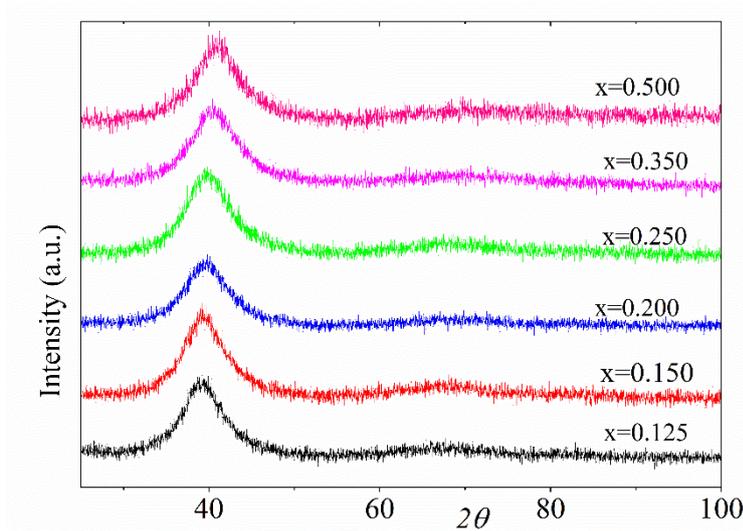

Figure 1. XRD traces of the $(TiZrNiCu)_{1-x}Ni_x$ alloys showing their amorphous state.

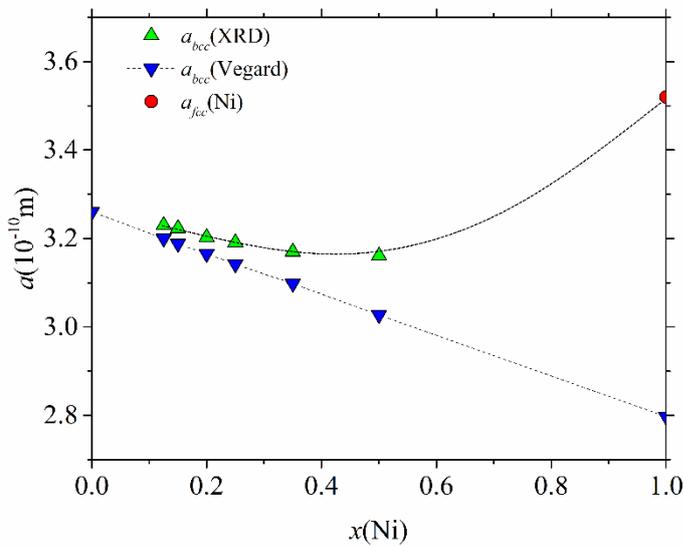

Figure 2. Lattice parameter *a* vs. concentration of Ni. Lines are guides for eye.

However, for $x \geq 0.25$ ($VEC \geq 7.0$) with increasing $x$ the experimental values of *a* progressively deviate from Vegard´s law for bcc-structure and for $x \rightarrow 1.0$ tend to extrapolate to that of pure fcc Ni. Thus, the analysis of XRD patterns of a-$(TiZrNbCu)_{1-x}Ni_x$ alloys seems to indicate a progressive change in local atomic arrangements from bcc-like ($x \leq 0.2$, $VEC \leq 6.8$) to possibly fcc-like. To our knowledge such a change in local atomic arrangements with composition (*VEC*) in the amorphous state of the same alloy system has not been reported so far. We note however that in binary metallic glasses of nickel with early transition metals there is a strong tendency towards chemical short range order (CSRO) [27,34]. This affects properties such as $N(E_F)$ and the Hall coefficient of those metallic glasses with elevated Ni-contents [34,35]. In order to elucidate this phenomenon, very recently we started to study the evolution of the radial distribution functions, *RDF(r)*, of our alloys with *x* at the Diamond synchrotron source [36]. First results for the average lattice parameter *a* of the alloys with $x \leq 0.25$ obtained from the positions of the first maxima of *RDF(r)* are in good agreement with the results in Fig. 2. and Table 1 in [21]. Regardless of its nature, a change in local atomic arrangements is expected to influence all the properties of a given alloy system. In what follows we show that this indeed happens in our alloys.

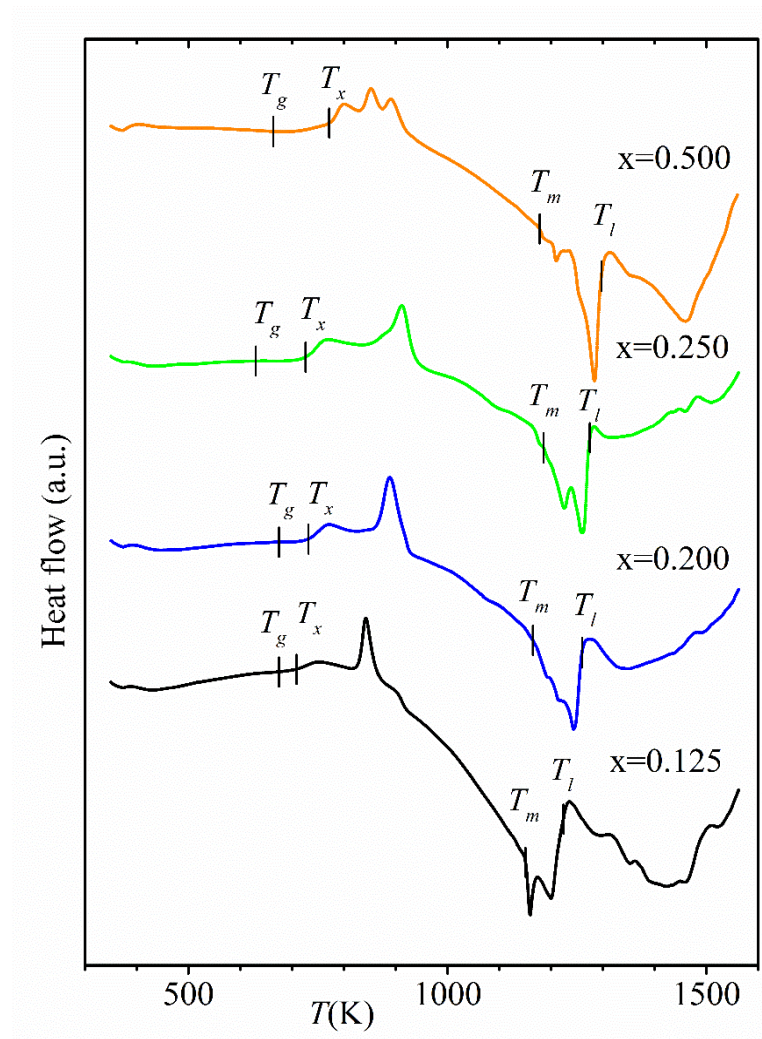

Figure 3. DSC/DTA data for selected $(TiZrNiCu)_{1-x}Ni_x$ alloys.

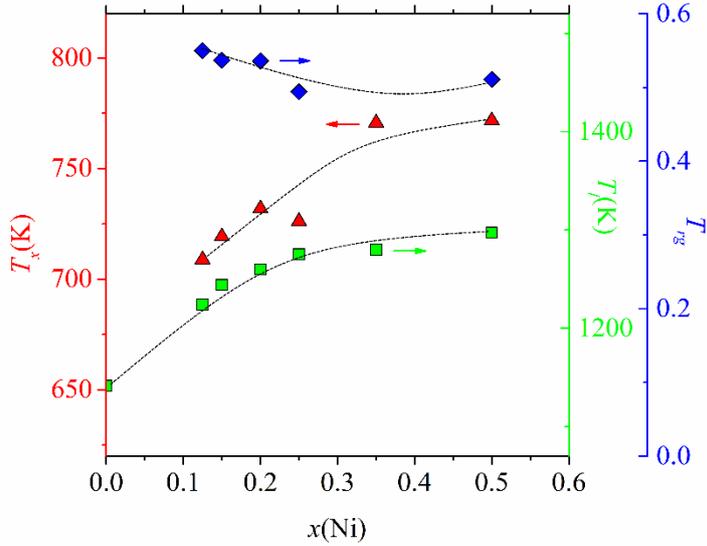

Figure 4. $T_x$ (left scale), $T_l$ (first right scale) and $T_{rg}=T_g/T_l$ (second right scale) vs. $x$. Lines are guides for eye.

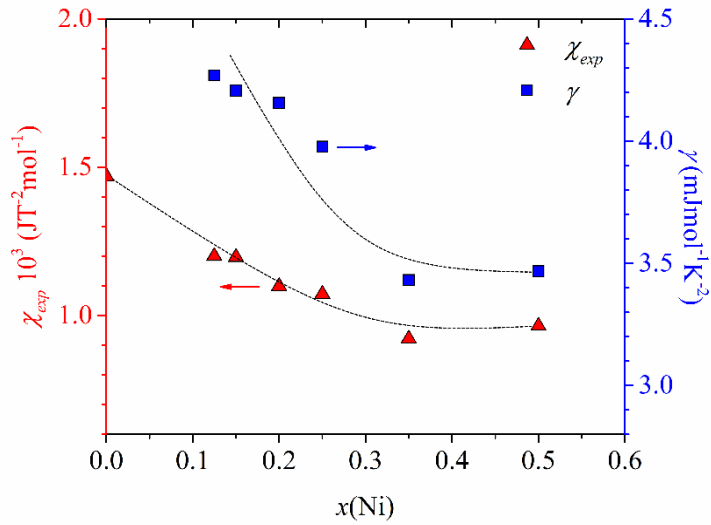

Figure 5. Room temperature magnetic susceptibilities $\chi_{exp}$ (left scale) and coefficient $\gamma$ from LTSH (right scale) vs. $x$. Lines are guides for eye.

Before discussing the properties of our a-HEAs which are more directly related to interatomic bonding and ES it is important to address the homogeneity of our samples. Indeed, quite often the distribution of constituent elements in HEAs is uneven, and this sometimes also occurs in HEAs showing single solid solution behaviour [7]. A detailed description of our SEM/EDS study for alloys with x≤0.25 has been previously reported [21]. Briefly, elemental mapping revealed random distribution of all constituents down to sub-micrometer scale and the elemental concentrations derived from EDS spectra agreed with nominal concentrations to within ≤ 2 at%, depending on selected element. The same type of SEM/EDS study performed on alloys with x = 0.35 and 0.50 revealed random distributions of all constituents, whereas their concentrations, derived from EDS spectra agreed with nominal ones to within ≤ 1 at% for all constituents. As illustrated in Fig. 3. DSC/DTA traces of our alloys showed well defined glass transition ($T_g$), crystallization ($T_x$), melting ($T_m$) and liquidus ($T_l$) temperatures. The exception was the alloy with $x = 0.35$ which showed a very smooth transition from glass to supercooled liquid which prevented a reliable estimate of its $T_g$. As noted earlier [21] our values of $T_g$ and $T_x$ for $x \leq 0.25$ agree quite well with those reported in [25]. As seen in Fig. 4., $T_x$ and $T_l$ which are associated with interatomic bonding increase with $x$, as is usual in a-TE-TL alloys [27,28,31,33]. However, in the present alloys the rate of increase of $T_x$ and $T_l$ with $x$ strongly decreases for $x \geq 0.25$ and they practically saturate beyond $x = 0.35$. The reduced glass transition temperature $T_{rg} = T_g/T_l$ of our alloys (which is an empirical criterion for glass forming ability (GFA)) shows little variation with $x$ and tends to saturate for $x \geq 0.25$ around $T_{rg} = 0.52$ which corresponds to modest GFA. Thus, all thermal parameters of our alloys show a clear change in dependencies on composition for $x \geq 0.25$. Since the local atomic arrangements determine the ES of the alloys (e.g. [20]) we expect a rather strong change in $N(E_F)$ thus also in the coefficient of electronic contribution to LTSH, $\gamma = \pi^2 k_B^2 N(E_F)/3$ of our alloys in going from a-HEAs ($x \leq 0.25$) to conventional TL-base amorphous alloys. Indeed, as seen from Fig. 5., in our alloys $\gamma$ ($N(E_F)$) values decrease on increasing $x$ as is usual in a-TE-TL alloys, but they saturate for $x \geq 0.35$, which does not occur in binary and ternary a-TE-TL alloys [27,31,32,34] at similar TL contents. As is well known, in spite of its complex origin [37] the magnetic susceptibility, $\chi_{exp}$, in a-TE-TL alloys varies with composition in qualitatively the same way as $\gamma$ [20,27]. (This similarity in variations of $\chi$ and $\gamma$ in binary and ternary a-TE-TL alloys probably reflects simple, approximately linear, concentration dependence of the orbital contribution to $\chi_{exp}$ [20,21,27,28,33].) Indeed as shown in Fig. 5. the variation of $\chi_{exp}$ at room temperature in our alloys with $x$ is qualitatively the same as that of $\gamma$, and $\chi_{exp}$ saturates (or reaches a shallow minimum) around $x = 0.35$. Thus, similar variations of $\gamma$ and $\chi_{exp}$ with composition seem to be a property of a-TE-TL alloys which does not depend on number of their components, and whether they are HEA or conventional alloys.

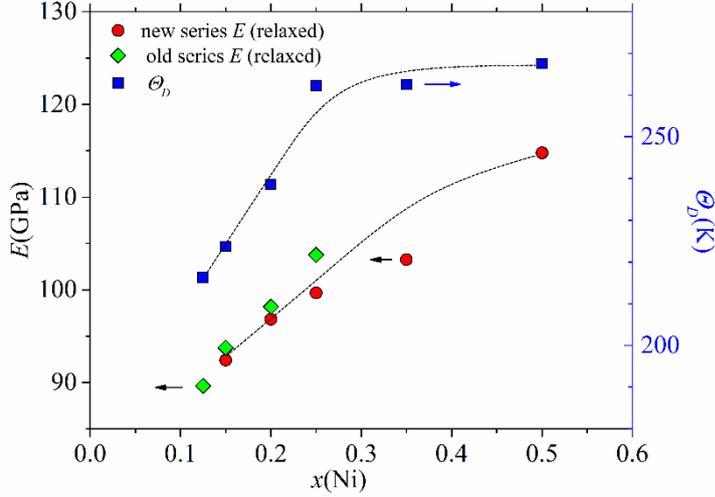

Figure 6. Youngs modulus (left scale) and $\theta_D$ (right scale) vs. $x$. Lines are guides for eye.

As noted earlier [27,31,38] there is a simple relationship between ES and the mechanical properties (including hardness) and thermal stability of a-TE-TL alloys. In particular, a decrease in $N(E_F)$ is accompanied by increases in Young´s modulus, Debye temperature and thermal stability (represented by $T_x$) in these alloys. Thus, a decrease of $N(E_F)$ reflects an increase in interatomic bonding, and accordingly the stiffness and parameters related to thermal stability and atomic vibrations increase, too. It is interesting to see whether this simple correlation (which is quite uncommon in crystalline metallic systems [31]) is still obeyed in the case of our alloys in which the local atomic arrangements probably change with Ni-content $x$ (Fig. 2.) or not. Fig. 4. shows that the correlation between $N(E_F)$ (Fig. 5.) and $T_x$ remained intact and Fig. 6. indicates that this may be true for $\theta_D$ and $E$, too. In particular, the increase of $\theta_D$ with $x$ practically stops for $x \geq 0.25$, whereas that of $E$ slows down beyond $x = 0.25$. We note the sensitivity of $E$ to quenched-in disorder and relaxation which introduces considerable uncertainty in the variation of $E$ with $x$ [21]. As already noted [21], RoM provides a poor description for $E$ and $\theta_D$ of our alloys.

4. Conclusion

The first comprehensive study of the transition from the regime of high entropy alloys (HEA) to that of conventional alloys based on late transition metals has been performed in amorphous (TiZrNbCu)$_{1-x}$Ni$_x$ alloy system. Careful analysis of the X-ray diffraction patterns seems to indicate that this transition is accompanied by a change in the local atomic arrangements which seem to change from approximately body centered cubic (bcc-like) for $x \leq 0.2$ to possibly approximately face centered cubic (fcc-like) for $x > 0.35$. Interestingly, this change occurs around the average valence electron number (*VEC*) about 7 which in single phase crystalline HEAs marks a border between pure bcc crystalline structure (*VEC* ≤ 7) and mixed bcc/fcc structure for *VEC* ≤ 8 [13]. The variations of all studied properties (thermal, mechanical, vibrational and magnetic) and of electronic structure (ES) (represented by the density of states at the Fermi level, $N(E_F)$, obtained from low temperature specific heat measurements) with $x$ support the change in local atomic structure. In particular, the Sommerfeld coefficient $\gamma$ (thus $N(E_F)$) becomes practically constant for $x \geq 0.35$ and the same occurs with the magnetic susceptibility. Similarly, the liquidus and the first crystallization temperatures (related to strength of interatomic bonding) as well as the Debye temperature (related to atomic vibrations) tend to saturate beyond $x = 0.25$. Further, the increase of Young´s modulus with $x$ also slows down for $x \geq 0.25$. The observed possibility of tuning the atomic and electronic structure within the amorphous state by changing the composition may be useful for fabrication of metallic glasses with predetermined properties.

Acknowledgement

We thank Prof. J.R. Cooper for useful comments and Dr. O. Lozada for the help with DSC/TGA. Our research was supported by the project "IZIP2016" of the Josip Juraj Strossmayer University of Osijek. I. A. Figueroa acknowledges the financial support of UNAM-DGAPA-PAPIIT, project No. IN101016.